\documentclass[sigconf]{acmart}
\usepackage{balance} 
\usepackage{xspace}
\usepackage{color}
\usepackage{caption}
\usepackage{subcaption}
\usepackage{tikz}
\usepackage{pgf}
\usepackage{pgfplots}
\pgfplotsset{compat=newest}
\usepgfplotslibrary{units,external,groupplots}
\usetikzlibrary{positioning}
\tikzsetexternalprefix{tikz/}
\usepackage{paralist}
\usepackage{pgfplotstable}
\usepackage{amsmath,amssymb}
\usepackage{wrapfig}
\usepackage{siunitx}
\usepackage[utf8]{inputenc}
\DeclareSIUnit\dBm{dBm}
\usetikzlibrary{patterns,angles,quotes,positioning,calc,shapes}
\usepackage{xfrac}

\graphicspath{{./img/}}

\usepackage{ifthen}
\newboolean{vspaceoptimization}
\setboolean{vspaceoptimization}{false}
\ifthenelse{\boolean{vspaceoptimization}}
{
\newcommand{\vsquish}[1]{\vspace{#1}}
}
{
\newcommand{\vsquish}[1]{}
}

\newboolean{authnotes}

\setboolean{authnotes}{false}

\ifthenelse{\boolean{authnotes}}
{

\newcommand{\fm}[1]{\footnote{{\bf\color{blue} FM: #1}}}
\newcommand{\mz}[1]{\footnote{{\bf\color{blue!50!orange} MZ: #1}}}
\newcommand{\db}[1]{\footnote{{\bf\color{red} DB: #1}}}
\newcommand{\st}[1]{\footnote{{\bf\color{red!70!yellow} ST: #1}}}
\newcommand{\rj}[1]{\footnote{{\bf\color{green!50!black} RJ: #1}}}
}
{
\newcommand{\fm}[1]{}
\newcommand{\db}[1]{}
\newcommand{\st}[1]{}
\newcommand{\mz}[1]{}
\newcommand{\rj}[1]{}
}

\newcommand\figref[1]{Figure~\ref{#1}}

\newcommand\secref[1]{Section~\ref{#1}}

\newcommand{\eg}{\emph{e.g.},\xspace}

\newcommand{\capt}[1]{\mdseries{\emph{#1}}}

\newcommand{\cps}{CPS\xspace}

\newcommand{\dBm}{\ensuremath{\,\text{dBm}}\xspace}

\author{Fabian Mager}
\authornote{Both authors contributed equally to this work.}
\affiliation{%
 \institution{TU Dresden}
}
\email{fabian.mager@tu-dresden.de}

\author{Dominik Baumann}
\authornotemark[1]
\affiliation{%
 \institution{MPI for Intelligent Systems}
}
\email{dominik.baumann@tuebingen.mpg.de}

\author{Romain Jacob}
\affiliation{%
 \institution{ETH Zurich}
}
\email{romain.jacob@tik.ee.ethz.ch}

\author{Lothar Thiele}
\affiliation{%
 \institution{ETH Zurich}
}
\email{thiele@ethz.ch}

\author{Sebastian Trimpe}
\affiliation{%
 \institution{MPI for Intelligent Systems}
}
\email{trimpe@is.mpg.de}

\author{Marco Zimmerling}
\affiliation{%
 \institution{TU Dresden}
}
\email{marco.zimmerling@tu-dresden.de}

\renewcommand{\shortauthors}{F.~Mager et al.}

\copyrightyear{2019}
\acmYear{2019}
\setcopyright{rightsretained}
\acmConference[IPSN '19]{The 18th International Conference on Information Processing in Sensor Networks (co-located with CPS-IoT Week 2019)}{April 16--18, 2019}{Montreal, QC, Canada}
\acmBooktitle{The 18th International Conference on Information Processing in Sensor Networks (co-located with CPS-IoT Week 2019) (IPSN '19), April 16--18, 2019, Montreal, QC, Canada}
\acmDOI{10.1145/3302506.3312483}
\acmISBN{978-1-4503-6284-9/19/04}

\begin{document}

\setdefaultleftmargin{2em}{}{}{}{}{}

\title[Fast Feedback Control and Coordination with Mode Changes]{Demo Abstract: Fast Feedback Control and Coordination with Mode Changes for Wireless Cyber-Physical Systems}


\begin{abstract}
This abstract describes the first public demonstration of feedback control and coordination of multiple physical systems over a dynamic multi-hop low-power wireless network with update intervals of tens of milliseconds.
Our running system can dynamically change between different sets of application tasks (\eg sensing, actuation, control) executing on the spatially distributed embedded devices, while closed-loop stability is provably guaranteed even across those so-called mode changes.
Moreover, any subset of the devices can move freely, which does not affect closed-loop stability and control performance as long as the wireless network remains connected.
\vspace{-1mm}
\end{abstract}



\maketitle


\vspace{-1mm}
\section{Introduction}
\label{sec:intro}
\vspace{-0mm}

Cyber-physical systems~(\cps) rely on embedded devices and wireless multi-hop networks to monitor and control physical systems at unprecedented scales, flexibility, and cost efficiency.
Realizing this potential for mission- or even safety-critical applications requires that the \emph{feedback loops} between sensors and actuators be closed quickly; for example, update intervals of tens of milliseconds are required to match the dynamics of mechanical systems (\eg for robot motion control and drone swarm coordination).
Moreover, because feedback control modifies the dynamics of physical systems, \emph{closed-loop stability} must be guaranteed despite notoriously unreliable wireless communication and dynamic changes in the configuration or behavior of the application.
For instance, the application may adapt at runtime to an external event by changing the set of tasks executed by certain devices, which is known as a \emph{mode change}.

Control over wireless and mode changes have been extensively studied (see~\cite{Chen2018,Zhang2013} for an overview).
The typical challenges include limited multi-hop throughput, varying communication delay, unpredictable message loss, and protocol-dependent constraints on what devices can exchange messages with one another.
However, most of the existing work lacks a validation on real platforms and networks, and none of the works that have been validated in practice consider \emph{fast} feedback control over \emph{multi-hop} networks.   

In our ICCPS 2019 paper,~\cite{Mager2019} we present the end-to-end design, formal analysis, and real-world validation of a wireless \cps that fills this gap.
As briefly outlined in \secref{sec:design}, we have extended the system to support applications that need to dynamically change between well-defined modes while ensuring closed-loop stability.
\secref{sec:scenario} describes the setup of our demo, what interested conference attendees can see at our booth, and how they can interact with the running system.
Overall, our demo showcases unprecedented functionality that is essential for emerging wireless \cps.%

\begin{figure*}
\begin{subfigure}[c]{0.415\linewidth}
    \centering
    \includegraphics[width=\linewidth]{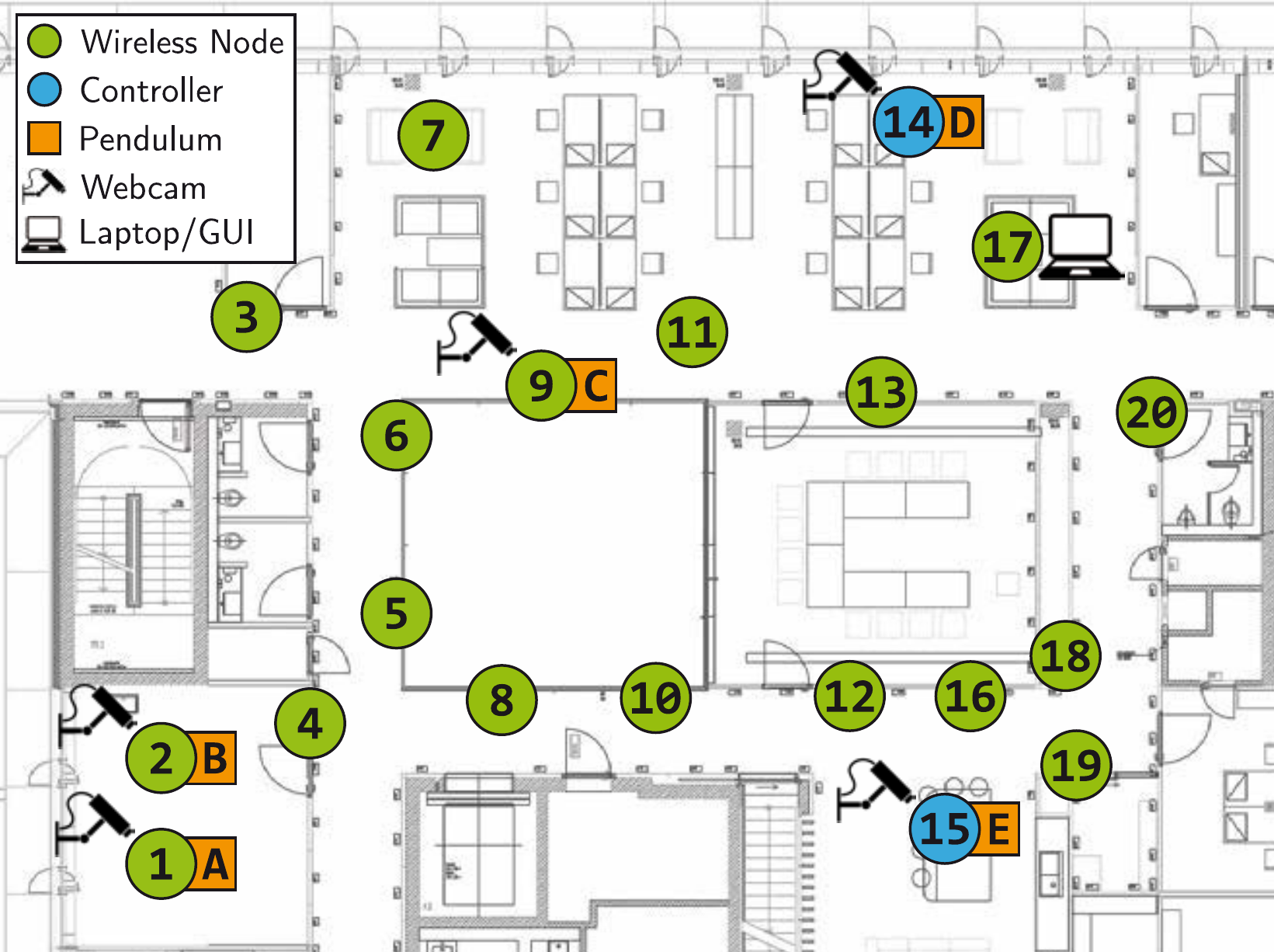}
    \subcaption{Envisioned layout of demo setup.}
    \label{fig:testbed}
\end{subfigure}
\hfill
\begin{subfigure}[c]{0.2\linewidth}
    \centering
    \includegraphics[width=\linewidth]{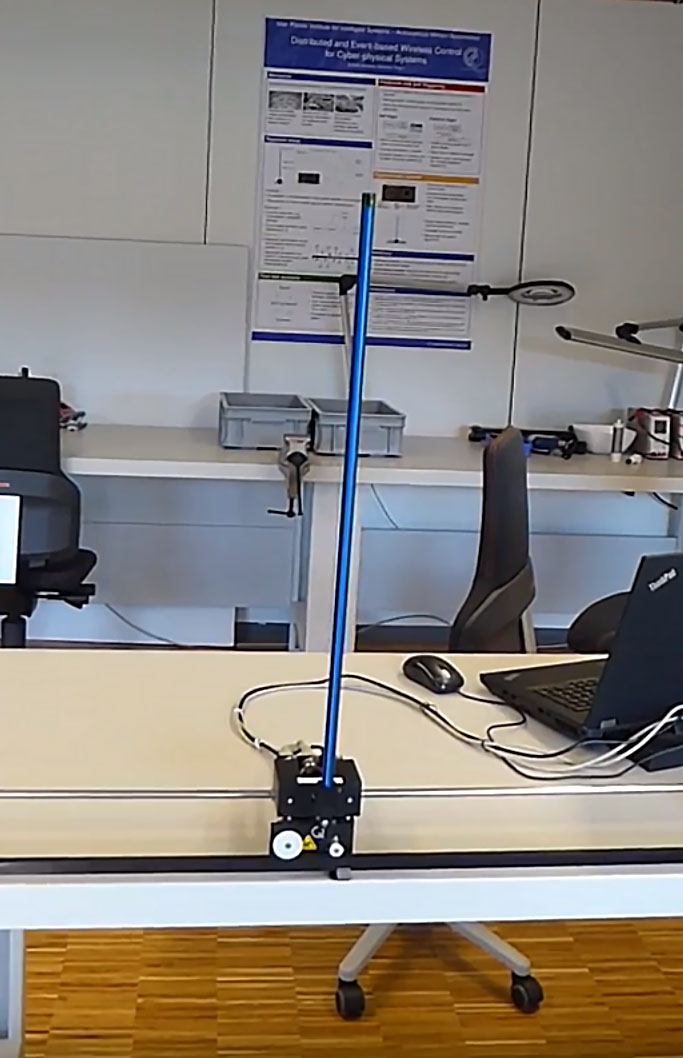}
    \subcaption{Inverted pendulum.}
    \label{fig:pendulum}
\end{subfigure}
\hfill
\begin{subfigure}[c]{0.357\linewidth}
    \includegraphics[width=\linewidth]{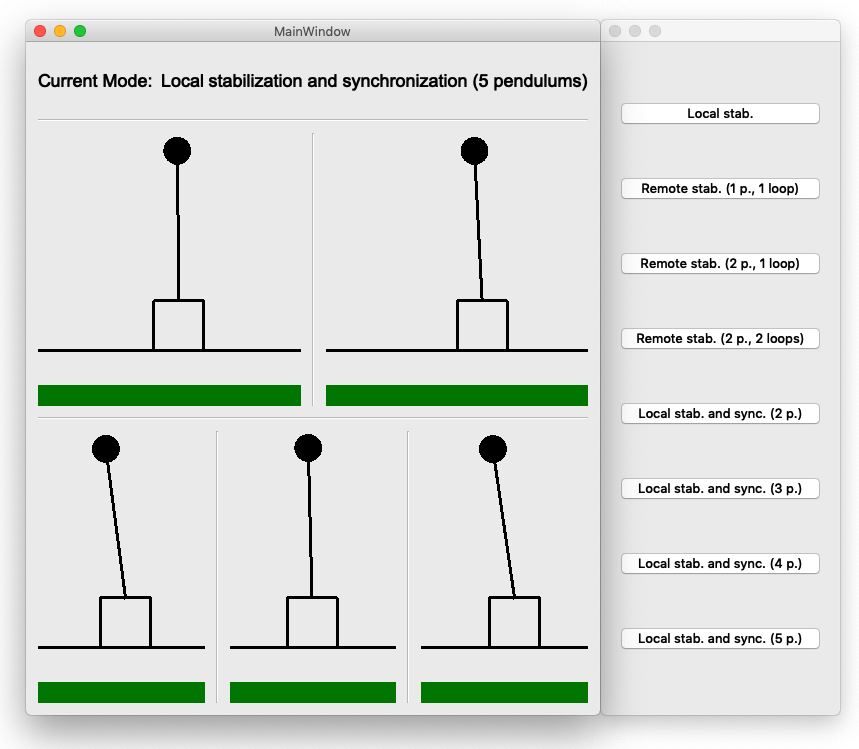}
    \subcaption{Graphical user interface.}
    \label{fig:gui}
\end{subfigure}
\vspace{-3mm}
\caption{Demo setup consisting of 20 wireless devices forming a three-hop network, two real and three simulated inverted pendulums, and a graphical user interface to request mode changes and to observe the state of all pendulums in realtime. \capt{Our system supports different control tasks, such as remote stabilization and synchronization of multiple inverted pendulums, by closing one or multiple feedback loops over the same low-power wireless multi-hop network at update intervals of tens of milliseconds. Interested conference attendees are invited to interactively change modes and carry around wireless devices to experience the flexibility and robustness of our system.}}
\vspace{-2mm}
\label{fig:overview}
\end{figure*}

\vspace{-0mm}
\section{System Overview}
\label{sec:design}
\vspace{-0mm}

To serve the needs mentioned above, we adopt a careful co-design approach whose tenet can be summarized as follows: Tackle the challenges through the design of the wireless embedded components (in terms of hardware and software) to the extent possible, and then consider the resulting key properties in the control design.

More concretely, we design a multi-hop low-power wireless protocol based on Glossy~\cite{Ferrari2011} that provides many-to-all communication with bounded end-to-end delay and accurate network-wide time synchronization.
Using a predictable dual-processor platform based on the Bolt interconnect,~\cite{Sutton2015} this protocol executes on a dedicated communication processor, while all application tasks (sensing, actuation, control) execute on a dedicated application processor.
We use the Time-Triggered Wireless~(TTW) framework~\cite{Jacob2018} to compute for each mode the corresponding schedule, ensuring that all timing requirements are met at minimum energy cost for wireless communication.
The schedules are distributed to the nodes before the system operation starts.
Using a novel mode-change protocol, nodes transition synchronously from one mode to another in a timely and safe manner, either in response to an event from the environment or in response to an event from within the system.
 
On the control side, the resulting key properties can be tackled by well-known techniques or safely neglected.
For example, we use state predictions to cope with communication delays and message loss, and neglect the worst-case jitter of \SI{\pm50}{\micro\second} on update interval and end-to-end delay as it is significantly smaller than those quantities (tens of milliseconds).
As a result, our solution is amenable to a formal analysis of all \cps components (wireless embedded, control, and physical systems), which we exploit to guarantee closed-loop stability for physical systems with linear time-in\-va\-ri\-ant~(LTI) dynamics in the presence of noise and mode changes.

\vspace{-1mm}
\section{Demo Setup and Scenario}
\label{sec:scenario}
\vspace{-0mm}

Depending on the conditions at the conference venue, we envision a demo setup as illustrated in \figref{fig:testbed}.
Based on a cyber-physical testbed we have built~\cite{Baumann2018} and used for the experiments in~\cite{Mager2019}, our demo setup consists of 20 battery-powered wireless devices forming a three-hop network.
The wireless network serves to close the feedback loops between one or multiple controllers and several physical systems.
We use inverted pendulums as physical systems, shown in \figref{fig:pendulum}, which have fast dynamics typical of real-world mechanical systems and thus require feedback with update intervals of tens of milliseconds.
The state of the pendulums is visualized in realtime via a graphical user interface (GUI), shown in \figref{fig:gui}, by interfacing a laptop to one of the wireless nodes.
Moreover, webcams provide a livestream of the real, spatially distributed pendulums at our demo booth.
The GUI also allows to dynamically request a change between different operating modes of our system.

In summary, interested conference attendees can observe and interact with our running system as follows:
\begin{compactitem}
	\item see stabilization and synchronization tasks over a low-power wireless multi-hop network at short update intervals in action;
	\item request changes between eight different well-defined modes we support via the GUI to test the runtime adaptability of our system while preserving closed-loop stability;
	\item carry around the wireless embedded devices to challenge the robustness of our system to significant network dynamics.\footnote{A video showing this part of our demo can be found at \url{https://youtu.be/19xPHjnobkY}.}
\end{compactitem}

\begin{acks}
We thank Harsoveet Singh, Felix Grimminger, and Katrin Kunz for their help with the
CPS testbed, and the TEC group at ETH Zurich for the design of the DPP platform and making it available to the public. We are also grateful to Quanser and made-for-science for generous support of our demonstration. This work was supported in part by the \grantsponsor{dfg}{German Research Foundation}{}~(DFG) within the Cluster of Excellence cfaed (grant \grantnum{dfg}{EXC 1056}), SPP 1914 (grants \grantnum{dfg}{ZI 1635/1-1} and \grantnum{dfg}{TR 1433/1-1}), and the Emmy Noether project NextIoT (grant \grantnum{dfg}{ZI 1635/2-1}), the Cyber Valley Initiative, and the Max Planck Society.
\end{acks}

\bibliographystyle{ACM-Reference-Format}
\bibliography{refs_short}

\end{document}